# Evaluation of carbon incorporation in sulfide thin films grown by hybrid pulsed laser deposition

Claire Wu[1], Mythili Surendran[1,2,*], Shin Muramoto[4], Alexander Grutter[5], Jayakanth Ravichandran[1,2,3]

1. Mork Family Department of Chemical Engineering and Materials Science, University of Southern California, Los Angeles, California 90089, USA
2. Core Center for Excellence in Nano Imaging, University of Southern California, Los Angeles, California 90089, USA
3. Ming Hsieh Department of Electrical and Computer Engineering, University of Southern California, Los Angeles, California 90089, USA
4. Material Measurement Laboratory, National Institute of Standards and Technology, Gaithersburg, MD, 20899, USA (ORCID 0000-0003-3135-375X)
5. NIST Center for Neutron Research, National Institute of Standards and Technology, Gaithersburg, MD, 20899, USA (ORCID 0000-0002-6876-7625)
*Currently at Lawrence Berkeley National Laboratory

Vapor-pressure mismatched materials such as transition metal chalcogenides have emerged as electronic, photonic, and quantum materials with scientific and technological importance. While hybrid pulsed laser deposition ($h$PLD) has emerged as the method of choice for epitaxial or textured growth of vapor-pressure mismatched materials, carbon (C) incorporation has been a persistent concern – especially in instances where organic chalcogen precursors are desired as a less hazardous alternative to more toxic chalcogen hydrides. However, the underlying mechanisms of such unintentional C incorporation and the effects on film growth and properties in $h$PLD are still elusive. Here, we report on the influence of C-containing side-products of organosulfur precursor pyrolysis in ZnS, BaTiS$_3$, and TiS$_2$ thin films grown by $h$PLD using a *tert*-butyl disulfide (TBDS) precursor. By combining structural characterization of X-ray diffraction (XRD) and atomic force microscopy (AFM) with secondary ion mass spectrometry (SIMS), we systematically investigate the role of temperature and TBDS partial pressures on film morphology and crystallinity. ZnS, TiS$_2$, and BaTiS$_3$ have optimal growth temperatures of 400°C, 500°C, and 700°C, respectively, and we observe that samples grown at temperatures above or below have increased C incorporation in the bulk and interface of the film, which correlates with poorer texture of the films as determined by XRD. On the other hand, highly textured films have minimal C at the film-substrate interface and within the film, which are comparable to films grown without TBDS in nominal vacuum conditions. We report TiS$_2$ growths with C incorporation dependent on TBDS growth pressure and determine that $10^{-1}$ Pa ($10^{-3}$ mbar) TBDS is the optimal growth pressure for minimal C contamination. At partial pressures greater than $10^{-1}$ Pa ($10^{-3}$ mbar) there is no preferential texture of the film, which is possibly caused by graphitization of the C, poisoning the interface and bulk of the film. This work opens opportunities for further understanding process-induced C impurity presence in $h$PLD grown thin film transition metal chalcogenides and chalcogenide perovskites and might have important implications for sulfide-based thin film technological applications.

# I. INTRODUCTION

Transition metal chalcogenides are a class of materials that are widely investigated for their desirable electronic, photonic, and electrochemical properties[1–3], with potential applications as transistors[4], sensors[5], photodetectors[6] and solar cells[7], and in electrocatalysis[8]. These materials tend to be vapor pressure mismatched, whose constituent elements have significantly different vapor pressures, especially sulfur containing compounds. This mismatch results in large differences in reactivity, surface adatom mobility, and sticking coefficients, leading to challenges in phase formation and crystallization of these materials. Specifically, controlling the anion stoichiometry is challenging due to competing reactions such as oxidation and the high propensity for sulfur vacancies at high growth temperatures.[9] $H_2S$ gas has been used as a sulfurizing agent and as a source of excess sulfur for thin film chalcogenide growth processes[10–12], however it is highly toxic, explosive, and corrosive[13], with low decomposition efficiency, leading to serious concerns for large scale applications. A less toxic replacement, *tert*-butyl disulfide (TBDS), $C_8H_{18}S_2$, was implemented as a sulfur precursor through the method of hybrid pulsed laser deposition (*h*PLD),[14] enabling efficient sulfur incorporation in the growth of epitaxial or quasi-epitaxial sulfide thin films. An environment of excess sulfur at relatively low temperatures is used to maintain sulfur stoichiometry at high vacuum conditions, while the ablation target controls the cationic stoichiometry. Although TBDS is a long-chained organo-disulfide that can undergo clean decomposition into highly volatile byproducts at low temperatures of (250 - 300) °C [15,16], carbon deposition at the film-substrate interface and within the bulk of the film can introduce carbon impurities due to pyrolysis side-products from organic ligands. Adventitious carbon can be assumed to be vaporized during the substrate heating process under ultra-high vacuum conditions at temperatures as low as 200 °C.[17] Carbon impurities have been reported to alter the optoelectronic properties of chalcogenide films,[18] such that uncontrolled impurities may lead to quenched luminescence,[19,20] or unintentional electronic doping and trap states that inhibit semiconducting behavior[21,22]. Thus, the presence of carbon impurity is an important metric of film quality that needs to be controlled and studied.

In this work, we investigate the role of carbon incorporation in the texturing and surface quality of thin film chalcogenides ZnS, $TiS_2$, and $BaTiS_3$. By employing secondary ion mass spectrometry, we systematically study carbon incorporation as a function of TBDS gas pressures and deposition temperature. X-ray diffraction and atomic force microscopy are used to relate film texture and surface roughness with carbon incorporation, and we report a systematic trend between synthesis temperature and carbon contamination for ZnS, $TiS_2$, and $BaTiS_3$. For each material, we reveal that the temperature at which the best texture is observed also has the lowest carbon contamination overall. Among the three materials, $TiS_2$ has the lowest carbon contamination at its optimal growth temperature. The relationship between synthesis temperature with bulk and interfacial carbon contamination in ZnS thin films is studied and reveals that interfacial carbon increases monotonically with temperature while bulk carbon has a minimum at a temperature, where the texture is maximized. Our work identifies the films' structural properties to be highly correlated with carbon incorporation, which underlines the importance of controlling and understanding carbon impurity levels during *h*PLD.

## II. EXPERIMENTAL

### A.  *Thin film deposition*

The thin films were grown by *h*PLD method using a 248 nm KrF excimer laser in a vacuum chamber (Demcon TSST) specifically designed for the growth of chalcogenides. The chamber was evacuated to a base pressure of ≈ $10^{-6}$ Pa ($10^{-8}$ mbar). The targets for ablation were densified using either sintering or cold isostatic pressing. High purity stoichiometric ZnS and $TiS_2$ (Alfa Aesar, 99.99%) powder was pressed into pellets with a diameter of ¾ inch and densified to >90% density by room temperature cold isostatic pressing. $BaTiS_3$ (BTS) powder was obtained through sulfurization of $BaTiO_3$ powder under $CS_2$ vapor at a flow rate of 1.67 x $10^{-7}$ $m^3$/s (10 sccm) for 24 h at 700°C. These dense pellets were used as the targets and were pre-ablated before growth. For all thin films, the organosulfur compound tert-butyl disulfide (TBDS) (99.999%, Sigma–Aldrich) was used as the sulfur precursor for the PLD growth. Since TBDS is a liquid with a relatively high vapor pressure at room temperature, it was introduced by thermal evaporation in a stainless-steel bubbler that was connected through a gas inlet system to the growth chamber. The bubbler was heated to a temperature of 130°C – 140°C using external heating tapes. The TBDS precursor delivery was controlled using a linear leak valve to achieve total chamber pressure of ≈ $10^{-1}$ Pa ($10^{-3}$ mbar). No carrier gas was used. Prior to deposition, $SrTiO_3$ (STO) substrates (Crystec GmbH) were sonicated in acetone and isopropyl alcohol. ZnS thin films were deposited via *h*PLD at temperatures in a range of 300 °C – 600 °C and a TBDS partial pressure of $10^{-1}$ Pa ($10^{-3}$ mbar) onto STO substrates. BTS thin films were deposited via *h*PLD at temperatures in a range of 400 °C – 700 °C and a TBDS partial pressure of $10^{-1}$ Pa ($10^{-3}$ mbar) onto STO substrates. $TiS_2$ thin films were deposited via *h*PLD at a temperature range of 400°C - 700°C and at a partial pressure range of $10^{-4}$ Pa to 5 x $10^{-1}$ Pa ($10^{-6}$ mbar - 5×$10^{-3}$ mbar) onto STO substrates. The fluence was fixed at 1.0 J

- 1.2 J cm$^{-2}$ and the target-substrate distance used was 75 mm. The films grown at a partial pressure of $10^{-1}$ Pa ($10^{-3}$ mbar) and above were cooled post growth at a rate of 10 °C min$^{-1}$ at an Ar-H$_2$S (4.97 molar ppm H$_2$S) partial pressure of $10^4$ Pa (100 mbar). The films grown at $10^{-4}$ Pa ($10^{-6}$ mbar) were cooled at a rate of 10 °C min$^{-1}$ under nominal vacuum pressure of $10^{-6}$ Pa ($10^{-8}$ mbar). All the films were characterized without any further processing such as annealing.

## B.  *Structural and Surface Characterization:*

The powder X-ray diffraction (XRD) studies were carried out on a Bruker D8 Advance diffractometer using a soller slit with Cu K$_\alpha$ ($\lambda$ = 1.5418 Å) radiation at room temperature. X-ray reflectometry (XRR) measurements were performed on the same diffractometer in a parallel beam geometry using a Göbel (parabolic) mirror setup. High resolution out-of-plane XRD measurements were performed on the same diffractometer using a Ge (004) two bounce monochromator with Cu K$\alpha$1 ($\lambda$=1.5406 Å) radiation at room temperature. Atomic force microscopy (AFM) measurements were performed on Bruker Multimode 8 atomic force microscope in peak force tapping mode with a ScanAsyst tip geometry to obtain the surface morphology and roughness.

## C.  *Secondary Ion Mass Spectrometry (SIMS) Characterization:*

The SIMS measurements were performed on an IONTOF IV (Münster, Germany) system using a time-of-flight analyzer for mass-to-charge ratio determination. The sputtering beam was a 20 keV Ar$^+_{2600\pm1000}$ cluster source, while the analysis beam was 30 keV Bi$_3^+$ liquid metal ion source. Sputtering and analysis were performed in a non-interlaced mode, with 2 analysis frames separated by between 2 – 10 sputtering frames, depending on film sputter rate and thickness. Ion dose per frame was in the range of $2.1 \times 10^{14}$ ions cm$^{-2}$ to $2.6 \times 10^{14}$ ions cm$^{-2}$ for the Ar cluster source and approximately $2 \times 10^{9}$ ions cm$^{-2}$ for the Bi$_3^+$. Sputter rate was also adjusted by varying the crater dimensions between (300 × 300) μm$^2$ to (600 × 600) μm$^2$. Negative and positive ions were measured in separate spots for each sample. The chamber base pressure was below $5 \times 10^{-7}$ Pa. To limit charging effects, the top surface of each sample was grounded to the sample holder using conductive tape.

Spectra were analyzed using SurfaceLab 7 to define a region of interest, perform mass calibrations, identify peaks with the appropriate compounds, and extract the total integrated peak intensity as a function of sputter time. The integrated peak intensity of each element was normalized in a point-by-point fashion to the integrated LMIG intensity during data collection. For comparison of peak intensities across samples, the intensity of each detected species was then normalized to the steady-state intensity of a selected substrate species, such as SrO (-) or Sr (+) (supplemental material Fig. S4). For the identification of film/substrate interface positions, some species are shown normalized to their maximum curve intensity. Depth is estimated based on the target film thickness and an assumption of a uniform sputter rate. In this work, we primarily discuss carbon content in terms of the $C_2^-$ peak intensity. However, we have examined the $C^+$, $C^-$, $C_2H^-$, $C_3^-$, and $C_4^-$ peaks for consistency and find general agreement in the trends across these species (supplemental material Fig. S5). On the other hand, some common hydrocarbon species such as CH$^-$, are isolated to the surface, suggesting a different origin.

## III. RESULTS AND DISCUSSION

Here we discuss the structure, surface quality, and carbon incorporation of ZnS, BTS and TiS$_2$ thin films grown at a range of temperatures and partial pressures of TBDS.

## A. Characterization of Zinc Sulfide Thin Films

In Fig. 1(a), we show representative X-ray diffraction patterns of ZnS thin films grown from 300°C – 600°C. The ZnS films show *c*-axis oriented texture as indicated by the 00*l* reflections in XRD scans in Fig. 1(a). At all temperatures outside of 400°C, we observed the degradation of the texture of ZnS films as shown by the $\theta$-$2\theta$ scans and the full-width-half-maximum values of the rocking curves shown in Fig. S1 in the supplemental material. The 00*l* peak at 400°C has thickness fringes, indicating smooth and well-defined interfaces compared to the films grown at other temperatures. As shown in Fig. 1(b) (i-iv), the AFM images indicate roughening of the films at temperatures above and below 400°C. The root-mean-squared roughness of the film was lowest when grown at 400°C at 0.67 ± 0.5 nm. The roughness of films grown at 300°C, 500°C, and 600°C are 1.3 ± 0.1 nm, 1.2 ± 0.1 nm, and 3.3 ± 0.3 nm, respectively. The SIMS intensity profile of the $C_2$ (-) ion in Fig. 1(c) indicates varying bulk and interface C contamination depending on the growth temperature. The $C_2$ (-) profiles are normalized by the SrO (–) SIMS profile at each given temperature and are shown in Fig. 1(d) as a reference. At 300°C and 500°C, there is significant carbon incorporation within the bulk of the film and a slight spike of carbon at the interface. At 400 °C, the optimal growth temperature, there is minimal carbon within the film and a low amount of carbon at the interface. At 600 °C, there is low carbon within the film, however, there is a high amount of carbon contamination at the film-substrate interface. Given the higher growth temperatures and oxygen deficient atmosphere, the interfacial carbon at 400°C – 600°C indicates the possibility of increased point defects due to an interfacial reaction between STO and TBDS. The film grown at 400°C has the lowest carbon concentration within the film and at the film-substrate interface, lowest roughness, and maximum texture. These suggest that an efficient cleavage of the carbon – tertbutyl group bond and the effective incorporation of sulfur into the film with minimal carbon give rise to an optimal growth temperature for textured growth. ZnS (−) and $S_2$ (−) SIMS profiles for all four films indicate uniform sulfur incorporation and are shown in supplemental material Fig. S3. At all temperatures, the usage of TBDS leads to carbon incorporation within the bulk of the ZnS film and the interface; however, the carbon is reduced to a minimum at the optimal growth temperature, which also seem to lead to the most textured film.

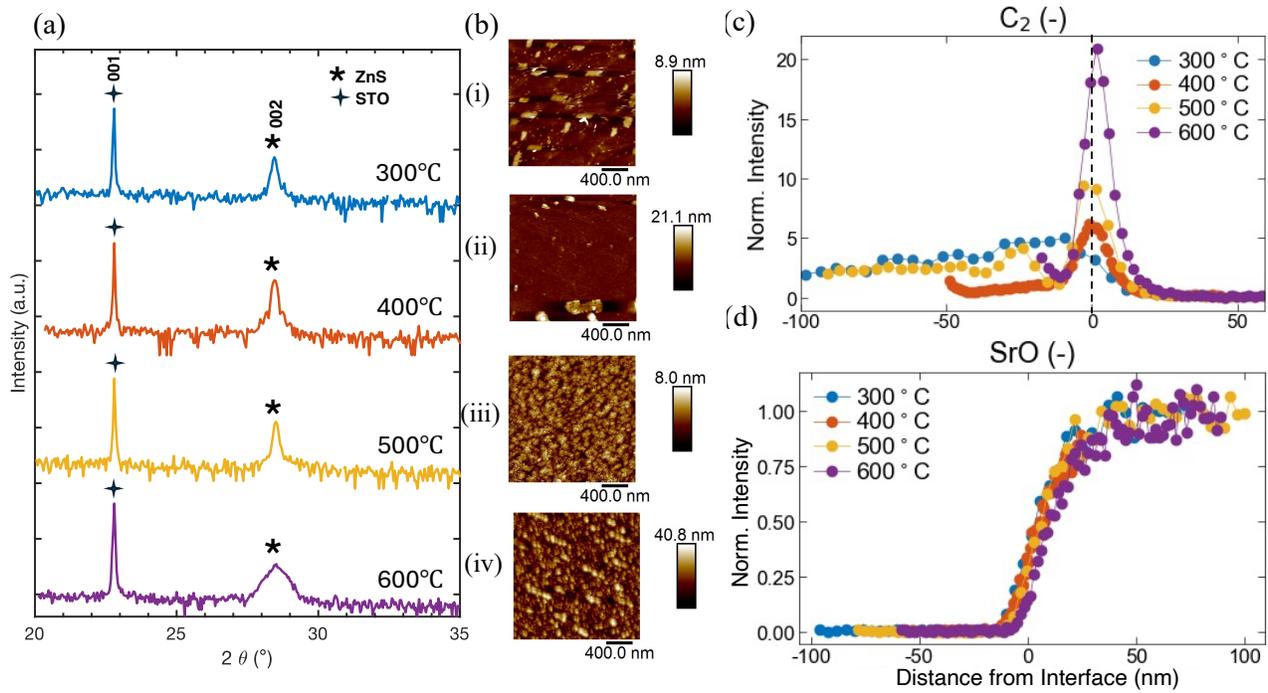

**FIG. 1**: (a) High resolution XRD $2\theta$-$\theta$ scans of ZnS thin films on SrTiO$_3$ substrate grown at temperatures 300°C - 600°C. (b) AFM images of (i) 300°C grown film with $R_q$=1.3 nm ± 0.08 nm (ii) 400°C grown film with $R_q$=0.7 nm ± 0.5 nm (iii) 500°C grown film with $R_q$=1.2 nm ± 0.1 nm (iv) 600°C grown film with $R_q$=3.3 ± 0.3 nm. (c) SIMS $C_2$ (-) profile of representative films grown from 300°C - 600°C with dotted line indicating film-substrate interface (d) SIMS SrO (-) ion profiles at temperatures 300°C - 600°C.

## B. Characterization of Barium Titanium Sulfide Thin Films

BTS thin films have significantly weaker texture than ZnS thin films, hence, to maximize the signal from the films, we used powder diffraction. In Fig. 2(a), we show representative X-ray diffraction patterns of BTS thin films grown from 400 °C – 700 °C. At 700°C, the BTS film shows 110 and 220 reflections in the XRD scan and at temperatures 400°C – 650°C, the 002 and 201 reflections appear, while the 110 reflection becomes weaker. There are additional substrate reflections in the XRD scan, including Bragg reflections at 41.87°, 44.52°, 44.87°, and 46.67°, corresponding to the 200 reflection of the STO substrate for Cu-K$\beta_1$, W-L$\alpha_1$, W-L$\alpha_2$, Cu-K$\alpha_2$, characteristic wavelengths, respectively. The 100 reflection of the STO substrate also exhibits the additional reflections associated with these wavelengths. The AFM images in Fig 2 (b) (i-iv) indicate the film roughening as temperature increases from 500°C to 700°C, with an RMS roughness of 1.0 ± 0.1 nm, 1.9 ± 0.3 nm, and 2.4 ± 0.3 nm, for films grown at 500°C, 650°C, and 700°C, respectively. In Fig 2 (c), the SIMS intensity profiles of the $C_2$ (-) ion show a non-monotonic behavior with temperature. At 700°C, the optimal temperature for texturing, there is low carbon concentration at the interface and bulk of the film. The film grown at 650°C has the largest amount of carbon concentration at the interface and low carbon within the bulk of the film. At 500°C, the carbon at the interface decreases and remains low within the bulk of the film. The SIMS intensity profile of $TiS_2$ (-) ion of BTS and SrO (-) ion from the substrate are shown in Fig. 2(d) as a reference, with the dotted line delineating the film-substrate interface. Like ZnS, the carbon incorporation is lowest in the bulk and the carbon content at the interface of the film drops to zero at the optimal growth temperature of BTS at 700°C. This points to carbon incorporation at the interface being texture-dependent and not simply temperature dependent.

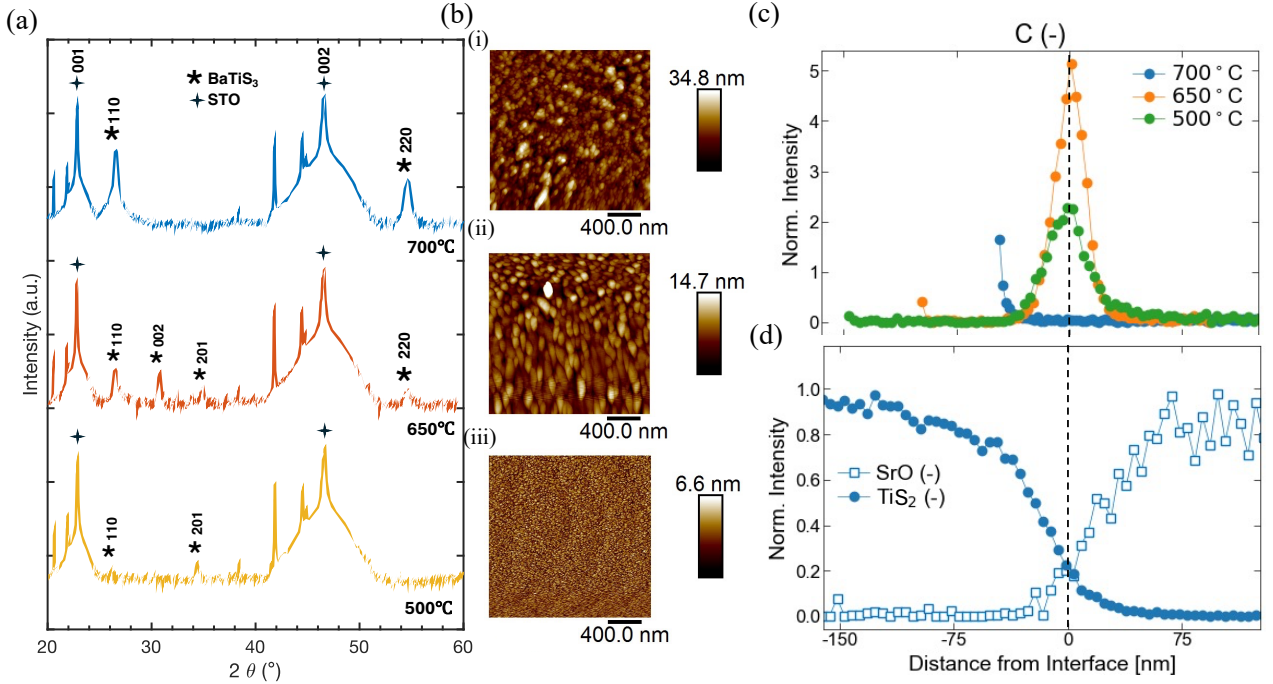

**FIG. 2:** (a) Powder diffraction $\theta$-$2\theta$ scans of BaTiS$_3$ on SrTiO$_3$ substrate grown at 500°C - 700°C. (b) AFM images of (i) 700°C grown film with $R_q$ = 2.4 ± 0.3 nm (ii) 650°C grown film with $R_q$ = 1.9 ± 0.3 nm (iii) 500°C grown film with $R_q$ = 1.0 ± 0.1 nm (c) SIMS carbon profile of representative films grown from 500°C - 700°C with dotted line indicating film-substrate interface (d) SIMS film and substrate ion profile of a 40 nm representative film grown at 700°C.

### C. *Characterization of Titanium Disulfide Thin Films*

In the case of TiS$_2$, we studied both the effect of temperature and the partial pressure of TBDS. First, we varied the growth temperature for the TiS$_2$ films, while the partial pressure of TBDS was kept at a constant value of $10^{-1}$ Pa ($10^{-3}$ mbar). We used powder diffraction and HRXRD to probe the texture depending on the strength of the signal obtained from the film. The powder diffraction shown in Fig. 3(a) reveals the dependence of film texture on growth temperature, with a 001 TiS$_2$ reflection from the film grown at 400°C and 500°C but no reflection when grown at 700°C. The TiS$_2$ 001 reflection at 400°C is shifted by $\Delta\theta$ = 0.55° from reported literature values of the TiS$_2$ 001 reflection, potentially indicating a non-stoichiometric film with larger *d*-spacing. The TiS$_2$ film grown at 500°C has the narrowest 2$\theta$ reflection, indicating higher crystalline quality and lower defect density. In some of scans with poorly textured films, we refrained from using filters or monochromators, and hence, there are additional substrate reflections in the XRD scan at 20.64°, 21.88°, 22.07°, and 22.85°, corresponding to the 100 STO reflection of Cu-K$\beta_1$, W-L$\alpha_1$, W-L$\alpha_2$, Cu-K$\alpha_2$ X-ray wavelengths, respectively. For the XRD scan of the film grown at 500°C, we used a Ni filter to eliminate the additional substrate reflections of STO. The texture of TiS$_2$ correlates with the relative C$_2$ (-) intensities shown in Fig. 3 (b), where the film grown at 500°C was found to have the highest texture and the least amount of carbon content at the interface and within the bulk. Relative to ZnS and BTS, TiS$_2$ has the lowest C$_2$ (-) intensity, which may be due to the layered 2D structure of TiS$_2$ compared to the 3D structures of BTS and ZnS. 2D materials have a lower likelihood of dangling bonds within the basal plane compared to a 3D material with an open crystal network. Additionally, intercalated carbon is unstable and has a higher formation energy compared to amorphous or graphitic clusters of carbon, which would prefer to segregate to the film-substrate interface.[23] The interfacial carbon seen at 400°C and 700°C agrees with a previous reports of MOCVD grown 2D material boron nitride and tungsten disulfide, which have the largest concentration of carbon at the film-substrate interface.[24,25] The AFM images in Fig. 3 (c) (i-iii) show a correlation between texture and roughness, with the film grown at 500°C exhibiting the highest texture and lowest roughness of 0.5 ± 0.1 nm while the films grown at 400°C and 700°C have a higher roughness of 2.4 ± 0.3 nm and 4.5 ± 0.1 nm, respectively. Thus, the optimal growth temperature for TiS$_2$ in this study is 500°C and the carbon contamination at the interface and bulk of the film is minimal with strong texturing of the film.

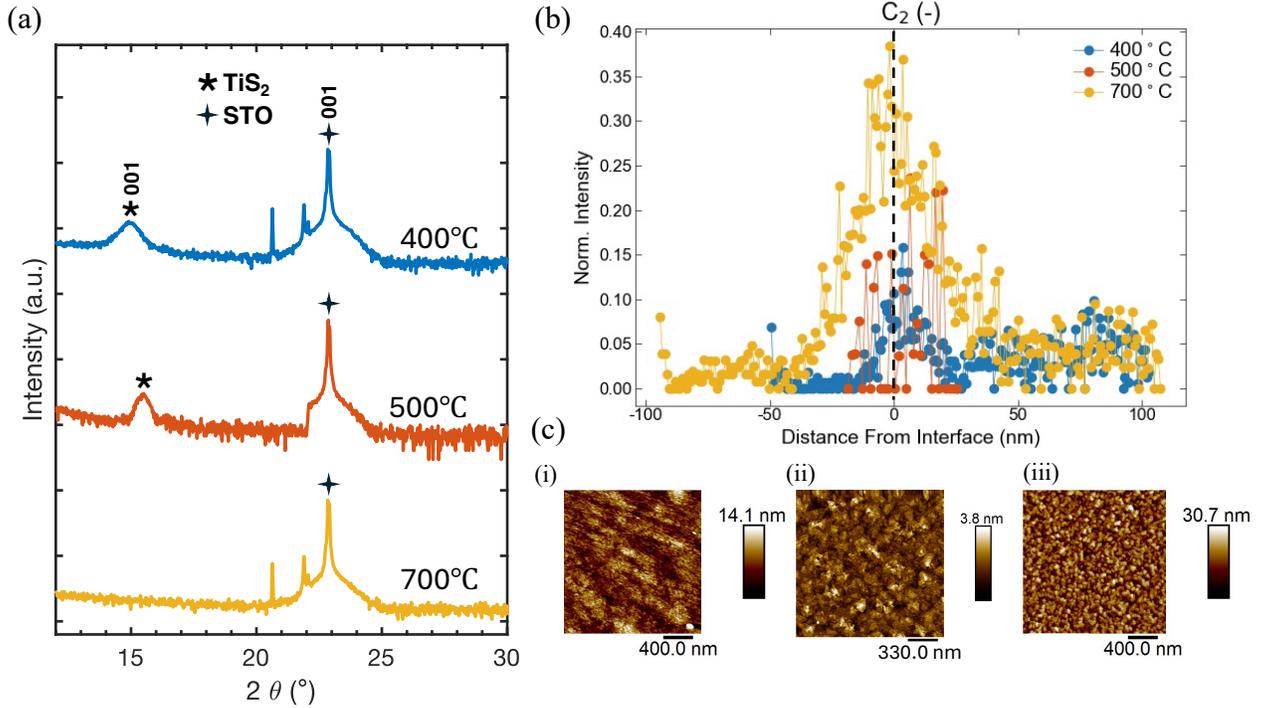

**FIG. 3**: (a) HR-XRD 2$\theta$-$\theta$ scans of TiS$_2$ on SrTiO$_3$ substrate grown at 500°C. (b) SIMS carbon profile of representative films grown from 400 – 700°C with dotted line indicating film-substrate interface (c) AFM images of (i) 400°C grown film with R$_q$=2.4 ± 0.3 nm (ii) 500°C grown film with R$_q$=0.5 ± 0.1 nm (iii) 700°C grown film with R$_q$=4.5 ± 0.1 nm.

Next, we kept the growth temperature as a constant at 500°C and varied the partial pressure of TBDS from $10^{-4}$ Pa to 5 x $10^{-1}$ Pa ($10^{-6}$ mbar to 5×$10^{-3}$ mbar) to study its effect on the texture of TiS$_2$ thin films. Under the nominal chamber base pressure of $10^{-4}$ Pa ($10^{-6}$ mbar) at 500°C, the TiS$_2$ was textured, exhibiting a 001 peak, in a manner similar to the optimal pressure of $10^{-1}$ Pa ($10^{-3}$ mbar) TBDS. However, the full-width-half-max (FWHM) of the rocking curve (supplemental material Fig. S2) at $10^{-4}$ Pa is 0.41°, much wider than the 0.05° found for the film grown at $10^{-1}$ Pa ($10^{-3}$ mbar) TBDS. Further, there was no preferential texture of the film grown at $5\times10^{-1}$ Pa ($5\times10^{-3}$ mbar), which is possibly caused by graphitization of the carbon, poisoning the interface and bulk of the film.[26] The texturing of the film was inversely correlated with the amount of carbon found in the SIMS analysis of the bulk of the film and at the film-substrate interface. As shown in Fig. 4 (b), at both $10^{-1}$ Pa ($10^{-3}$ mbar) and $10^{-4}$ Pa ($10^{-6}$ mbar), carbon incorporation is minimal at both locations, but this becomes relatively high at $5\times10^{-1}$ Pa ($5\times10^{-3}$ mbar). Our results also suggest that the pressure of $5\times10^{-1}$ Pa TBDS may be sufficiently high to cause extended diffusion of carbon species into the substrate, as non-trivial amount of carbon was detected inside the STO substrate. Hence, we conclude that this partial pressure of TBDS may be too high for the growth of $TiS_2$ thin films. In this study, we note an optimal growth pressure of $10^{-1}$ Pa ($10^{-3}$ mbar) mbar for $TiS_2$ at which the films showed the strongest texture and the lowest overall incorporation of carbon.

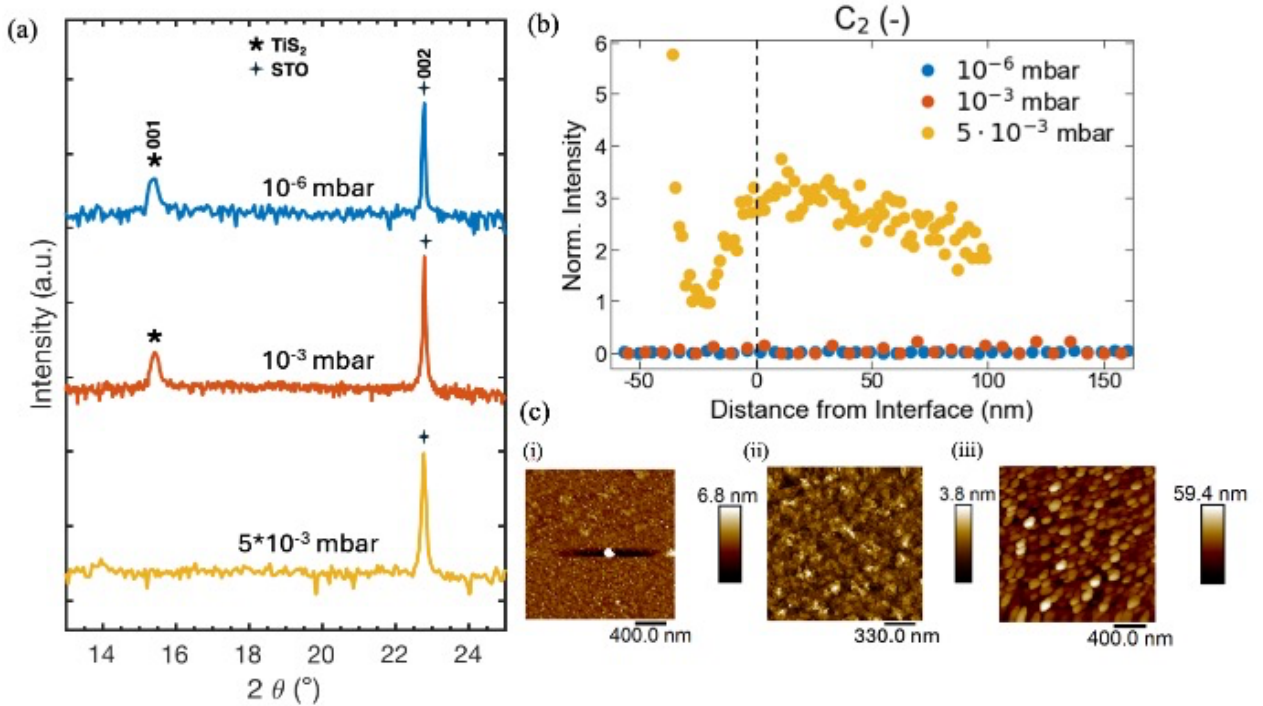

**FIG. 4**: (a) HR-XRD $2\theta$-$\theta$ scans of $TiS_2$ on $SrTiO_3$ substrate grown at 500°C. (b) SIMS $C_2$ (-) profile of representative films grown at 500°C at pressures of $5\times10^{-1}$ Pa to $10^{-4}$ Pa ($10^{-6}$

mbar to $5\times10^{-3}$ mbar) with the dotted line indicating the film-substrate interface position (c) AFM images of (i) $10^{-6}$ mbar grown film with $R_q=1.2 \pm 0.1$ nm (ii) $10^{-3}$ mbar grown film with $R_q=0.5 \pm 0.1$ nm (iii) $5\times10^{-3}$ mbar grown film with $R_q=9.1 \pm 0.8$ nm.

### D. SIMS Analysis

To further understand the incorporation of carbon and its dependence on temperature and partial pressure, $C_2$ (-) ion profiles were integrated and the areas compared. The $C_2$(-) intensity was normalized by the substrate signal SrO (-) to account for sample-to-sample intensity variations, and the x-axis was scaled to the sputtered depth and the film thickness placed at the interface position to account for variations in film thickness and sputtering rate. The 'interface' and 'bulk' position in the profile are delineated by changes in slope of the SIMS curves of Sr (+) and ZnS (+) at each respective temperature as shown in Fig 5(b). The integrated $C_2$ (-) curves for the bulk region of the ZnS, BTS, and $TiS_2$ films are shown in Fig. 5 (a). The bulk carbon content is normalized by film thickness to account for thickness variations between samples. ZnS and BTS have a clear minimum at their optimal growth temperatures of 400°C and 700°C, respectively. ZnS has a trend in which the extent of carbon incorporation increases as its growth temperature deviates in both directions from its optimal growth temperature (the difference from the optimal temperature is noted as $\Delta T_{OPT}$). For BTS, while the films do not grow textured under *h*PLD conditions above 700°C, the extent of carbon incorporation increases as the temperature is reduced, similar to the ZnS film. $TiS_2$ has low bulk carbon incorporation across all temperatures, indicating that the 2-D layered structure of $TiS_2$ has a lower likelihood of carbon entering the interstitial sites compared to the 3-D structures of ZnS and BTS. In addition to studying the bulk carbon in all three materials, carbon content at the interface vs. in the bulk of ZnS films was further investigated, as shown in Figure 5 (b). The interfacial carbon incorporation increases monotonically with growth temperature, which may be due to increased diffusion of graphitic carbon towards the film-substrate interface. As shown, both the film chemistry and growth temperatures are seen to affect the extent of carbon incorporation into the bulk and interfacial regions, with differences also seen in the amount of carbon incorporated into these regions.

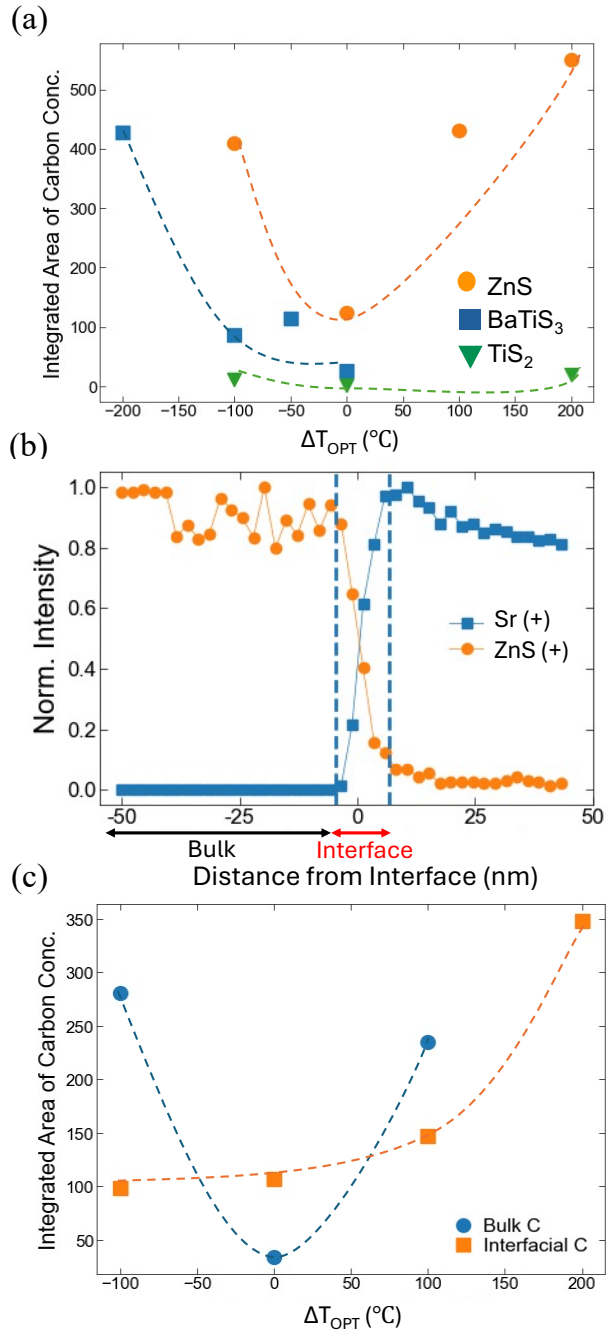

**FIG. 5:** (a) Integrated area of the $C_2$ (-) ion profile for the bulk region of the $BaTiS_3$, ZnS, and $TiS_2$ films at growth temperatures of 300°C - 700°C, with $\Delta T_{OPT}$ (difference from the optimal growth temperatures of 700°C, 400°C, and 500°C, respectively). (b) Representative SIMS profile delineating how the bulk and interface regions are determined for ZnS film grown at 400°C. (c) Bulk and interfacial $C_2$ (-) ion integrated area of ZnS films at 300°C - 600°C. Dotted lines are added as guides to the eye.

## IV. SUMMARY AND CONCLUSIONS

We report an evaluation of C incorporation using SIMS in thin film sulfide growth using TBDS as an organosulfur precursor in $h$PLD. ZnS, TiS$_2$, and BTS films have optimal growth temperatures of 400°C, 500°C, and 700°C, respectively, at which carbon incorporation within the bulk region of the film was lowest. Although our analysis was semi-quantitative, TiS$_2$ had the lowest C contamination at its optimal growth temperature among the three materials investigated in this study. The TiS$_2$ films showed maximal texture at an optimal value of $10^{-1}$ Pa ($10^{-3}$ mbar) for the partial pressure of TBDS. In all these cases, we found that there is an inverse correlation between improvement of film texture and the concentration of carbon incorporation into the bulk region of the film, with a clear trend for the ZnS and BTS films that points to an optimal growth temperature having a minimum carbon incorporation. Specifically, for the ZnS film, it was seen that the carbon incorporation in the bulk region of the film would increase outside of the optimal growth temperature, but the carbon incorporation at the film-substrate interface increased monotonically with temperature, indicating the possibility of increased point defects due to an interfacial reaction between TBDS and the film-substrate interface. Non-thermal cleavage of the organosulfur precursors could pave way to broadened growth window in terms of temperature for chalcogenide thin films using $h$PLD. Our findings elucidate the impact of carbon incorporation on the quality of thin film sulfides and can likely be applied to other thin film growth processes that use TBDS as an organosulfur precursor.

# SUPPLEMENTARY MATERIAL

TABLE I. Zinc sulfide film thicknesses and growth temperatures. All of the zinc sulfide films have been deposited on SrTiO$_3$ substrates at a partial pressure of $10^{-1}$ Pa ($10^{-3}$ mbar) of TBDS at a fluence of (0.95 - 1.1) J cm$^{-2}$, a repetition rate of 5 Hz, and cooled at 10 Pa ($10^{-1}$ mbar) under Ar-H$_2$S gas.

| Film No. | 1 | 2 | 3 | 4 |
|---|---|---|---|---|
| **Thickness (nm)** | 110 ± 2 | 46 ± 1 | 110 ± 2 | 55 ± 1 |
| **Temperature (°C)** | 300 | 400 | 500 | 600 |

TABLE II. Barium titanium sulfide film thicknesses, fluences and growth temperatures. All of the BaTiS$_3$ films have been deposited on SrTiO$_3$ substrates at $10^{-1}$ Pa ($10^{-3}$ mbar) under TBDS gas and at a repetition rate of 2 Hz, then cooled at 10 Pa ($10^{-1}$ mbar) under Ar-H$_2$S gas.

| Film No. | 5 | 6 | 7 | 8 |
|---|---|---|---|---|
| **Thickness (nm)** | 167 ± 2 | 125 ± 2 | 71 ± 1 | 40 ± 1 |
| **Fluence (J cm$^{-2}$)** | 1.25 | 1.25 | 1 | 1 |
| **Temperature (°C)** | 400 | 500 | 650 | 700 |

TABLE III. Titanium sulfide film thicknesses, growth pressures and growth temperatures. All of the TiS$_2$ films have been deposited on SrTiO$_3$ substrates at a repetition rate of 5 Hz and fluence of (1.0 - 1.1) J cm$^{-2}$, then cooled at 10 Pa ($10^{-1}$ mbar) under Ar-H$_2$S gas.

| Film No. | 11 | 12 | 13 | 14 | 15 |
|---|---|---|---|---|---|
| **Thickness (nm)** | 95 ± 1 | 36 ± 1 | 40 ± 1 | 26 ± 1 | 90 ± 1 |
| **Pressure (Pa)** | $10^{-1}$ | $10^{-4}$ | $10^{-1}$ | 5 x $10^{-1}$ | $10^{-1}$ |

| Temperature (°C) | 400 | 500 | 500 | 500 | 700 |

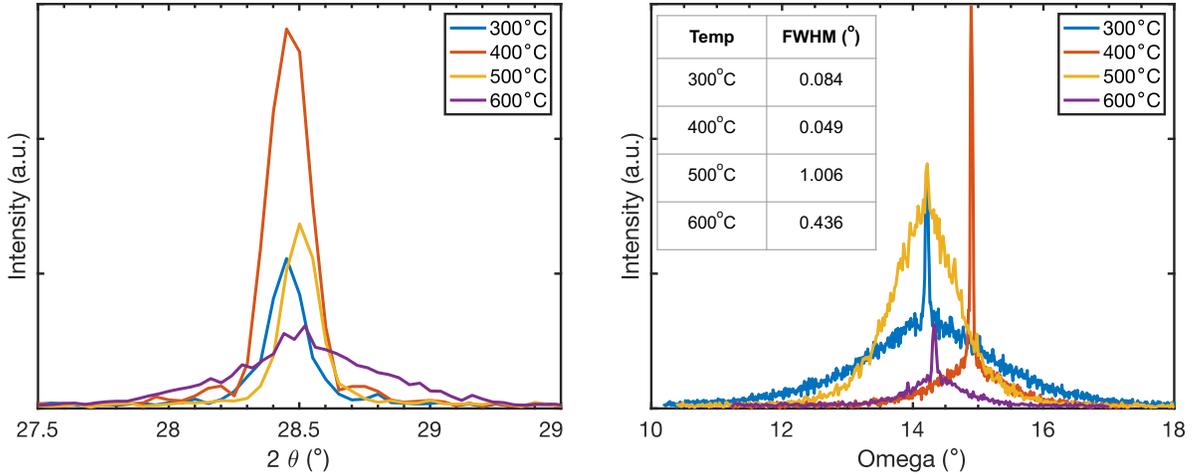

**FIG S1. (a)** HRXRD scans and **(b)** rocking curve scans showing textured ZnS films grown at temperatures in a range of (300 - 600) °C. Rocking curve scans show broadening of FWHM at temperatures outside of the optimal temperature of 400°C.

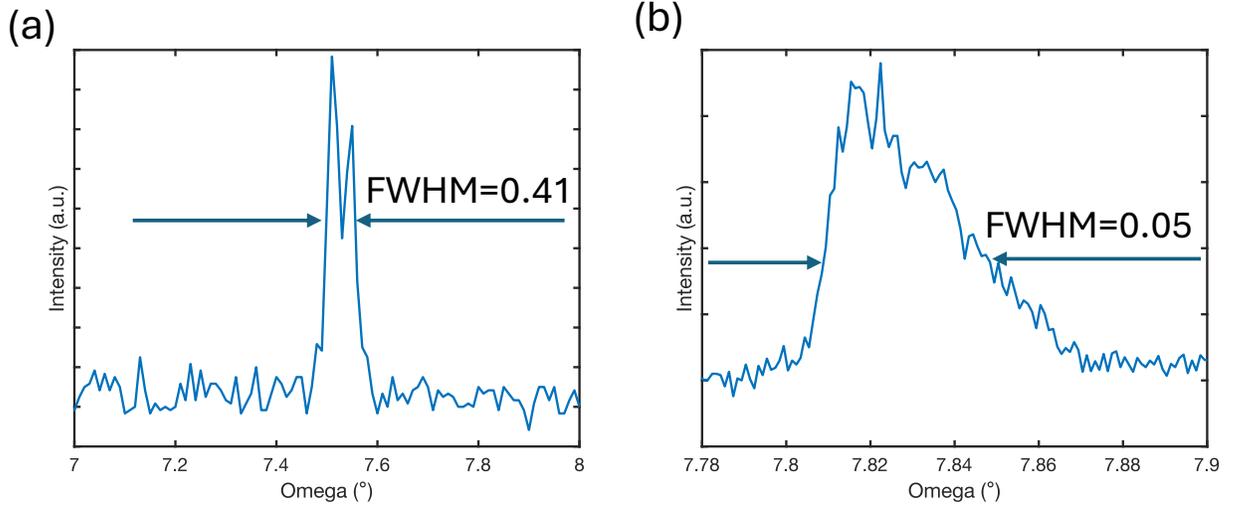

**Fig. S2.** TiS$_2$ rocking curve for film grown at 500°C at **(a)** $10^{-4}$ Pa ($10^{-6}$ mbar) and **(b)** $10^{-1}$ Pa ($10^{-3}$ mbar) TBDS.

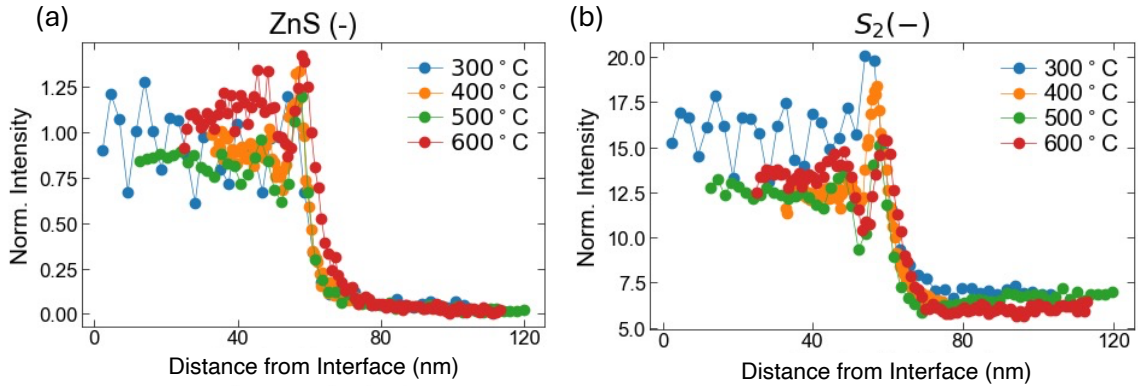

**Fig. S3: (a)** ZnS (-) and **(b)** $S_2$ (-) SIMS profiles of ZnS films on SrTiO$_3$ grown at 300 – 600 °C and $10^{-1}$ Pa.

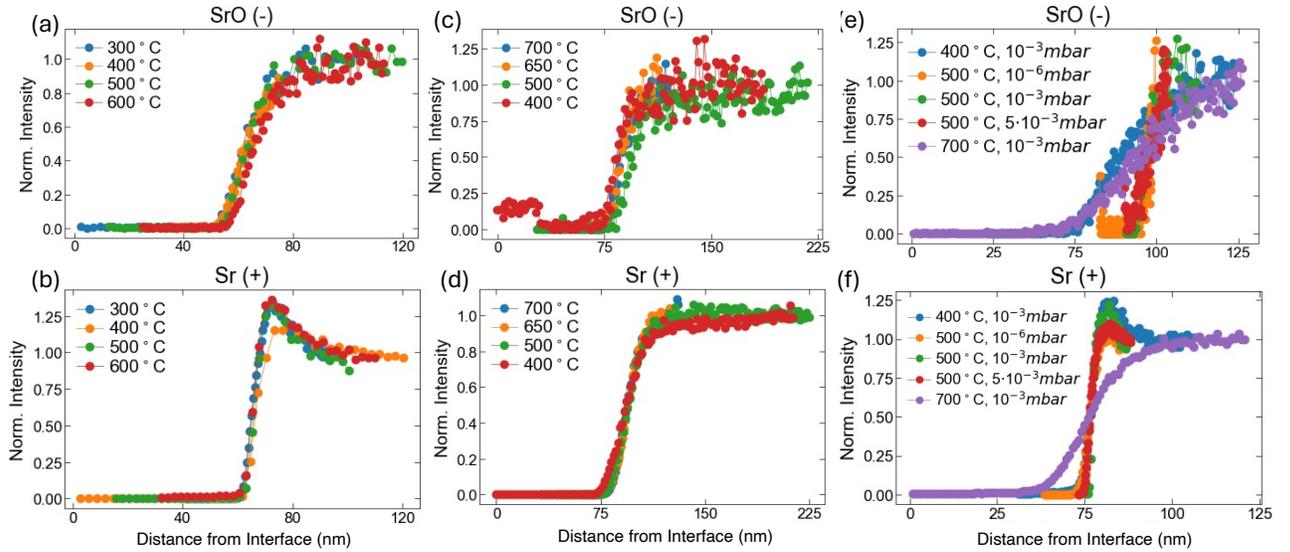

**Fig. S4: (a)** SrO (-) and **(b)** Sr (+) SIMS profiles for ZnS films grown at 300 – 600 °C. **(c)** SrO (-) and **(d)** Sr (+) SIMS profiles for BaTiS$_3$ films grown at 400 – 700 °C **(e)** SrO (-) and **(f)** Sr (+) SIMS profiles for TiS$_2$ films grown at 400 – 700 °C and pressure ranges from $10^{-4}$ - 5 x $10^{-1}$ Pa ($10^{-6}$ - 5•$10^{-3}$ mbar).

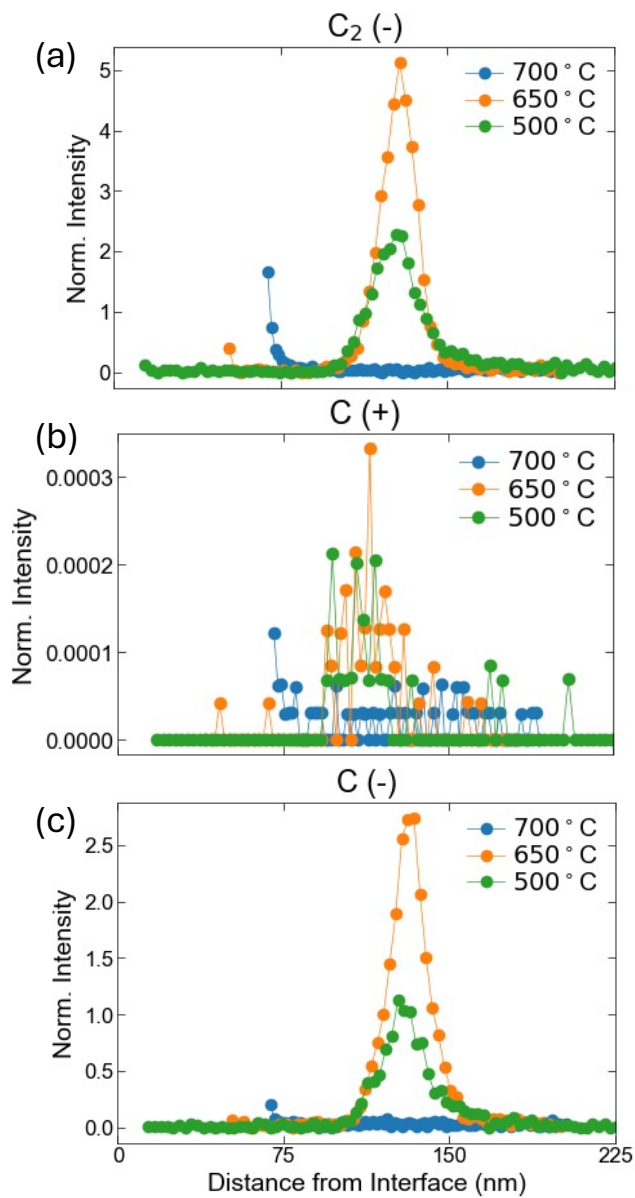

**Fig S5: (a)** $C_2$ (-) **(b)** C (+) and **(c)** C (-) SIMS profiles for $BaTiS_3$ films grown from 500 – 700 °C and $10^{-1}$ Pa ($10^{-3}$ mbar).

# ACKNOWLEDGMENTS

This work in part was supported by Army Research Office award no. W911NF-24-1-0164. C. Wu acknowledges the National Science Foundation of the United States under grant number DGE-1842487. The modification to the growth system to enable hybrid PLD (*h*PLD) was supported by an Air Force Office of Scientific Research grant no. FA9550-22-1-0117. The authors gratefully acknowledge the use of facilities at the Core Center for Excellence in Nano Imaging at University of Southern California for the results reported in this manuscript. Certain commercial equipment, instruments, or materials, commercial or non-commercial, are identified in this paper in order to specify the experimental procedure adequately. Such identification does not imply recommendation or endorsement of any product or service by NIST, nor does it imply that the materials or equipment identified are necessarily the best available for the purpose. Unless otherwise noted, NIST work was funded solely by the United States Government.

## AUTHOR DECLARATIONS

**Conflicts of Interest** *(required)*

The authors have no conflicts to disclose.

**Author Contributions** *(if applicable)*

C. Wu, A. Grutter and J. Ravichandran conceived the project. C. Wu and M. Surendran grew the thin films. A. Grutter and S. Muramoto performed SIMS measurements. J. Ravichandran supervised the project. C. Wu wrote the manuscript with input from A. Grutter and J. Ravichandran. All authors contributed to the revision of the manuscript.

## DATA AVAILABILITY

The data that support the findings of this study are available from the corresponding author upon reasonable request.